\documentclass[acmsmall,screen]{acmart}

\usepackage{xcolor,colortbl}
\usepackage{graphicx}
\usepackage{color}
\usepackage{pifont}
\usepackage{ifthen}
\usepackage{multirow}
\usepackage{booktabs}
\usepackage[inline]{enumitem}
\usepackage[ruled,linesnumbered]{algorithm2e}
\usepackage{tcolorbox}
\usepackage{booktabs}
\usepackage{tabularx}
\newcolumntype{L}{>{\raggedright\arraybackslash}X}%
\newcolumntype{R}{>{\raggedleft\arraybackslash}X}%
\newcolumntype{C}{>{\centering\arraybackslash}X}%
\usepackage{listings}
\usepackage[pdf]{graphviz}

\hypersetup{
	breaklinks=true
}

\usepackage{relsize}
\usepackage{threeparttable}
\usepackage{amsmath}
\usepackage{mathtools}
\usepackage{tikz}
\usepackage{fontawesome}
\usepackage{caption}
\usepackage{subcaption}
\usepackage{titlesec}
\usepackage{xspace}
\usetikzlibrary{shapes,positioning}
\makeatletter
\let\old@lstKV@SwitchCases\lstKV@SwitchCases
\def\lstKV@SwitchCases#1#2#3{}
\makeatother
\usepackage{lstlinebgrd}
\makeatletter
\let\lstKV@SwitchCases\old@lstKV@SwitchCases

\lst@Key{numbers}{none}{%
	\def\lst@PlaceNumber{\lst@linebgrd}%
	\lstKV@SwitchCases{#1}%
	{none:\\%
		left:\def\lst@PlaceNumber{\llap{\normalfont
				\lst@numberstyle{\thelstnumber}\kern\lst@numbersep}\lst@linebgrd}\\%
		right:\def\lst@PlaceNumber{\rlap{\normalfont
				\kern\linewidth \kern\lst@numbersep
				\lst@numberstyle{\thelstnumber}}\lst@linebgrd}%
	}{\PackageError{Listings}{Numbers #1 unknown}\@ehc}}
\makeatother

\title[Understanding and Characterizing Mock Assertions in Unit Tests]{Understanding and Characterizing Mock Assertions \\ in Unit Tests}



\setcopyright{cc}
\setcctype{by}
\acmDOI{10.1145/3715741}
\acmYear{2025}
\acmJournal{PACMSE}
\acmVolume{2}
\acmNumber{FSE}
\acmArticle{FSE026}
\acmMonth{7}
\received{2024-09-13}
\received[accepted]{2025-01-14}

\author{Hengcheng Zhu}
\orcid{0000-0002-3082-5957}
\email{hzhuaq@connect.ust.hk}
\affiliation{%
	\institution{The Hong Kong University of Science and Technology}
	\city{Hong Kong}
	\country{China}
}

\author{Valerio Terragni}
\orcid{0000-0001-5885-9297}
\email{v.terragni@auckland.ac.nz}
\affiliation{%
	\institution{The University of Auckland}
	\city{Auckland}
	\country{New Zealand}
}

\author{Lili Wei}
\orcid{0000-0002-2428-4111}
\email{lili.wei@mcgill.ca}
\affiliation{%
	\institution{McGill University}
	\city{Montréal}
	\country{Canada}
}

\author{Shing-Chi Cheung}
\authornote{Shing-Chi Cheung is the corresponding author.}
\orcid{0000-0002-3508-7172}
\email{scc@cse.ust.hk}
\affiliation{%
	\institution{The Hong Kong University of Science and Technology}
	\city{Hong Kong}
	\country{China}
}

\author{Jiarong Wu}
\orcid{0000-0001-6126-303X}
\email{jwubf@cse.ust.hk}
\affiliation{%
	\institution{The Hong Kong University of Science and Technology}
	\city{Hong Kong}
	\country{China}
}

\author{Yepang Liu}
\orcid{0000-0001-8147-8126}
\email{liuyp1@sustech.edu.cn}
\affiliation{%
	\department{Department of Computer Science and Engineering}
	\department{Research Institute of Trustworthy Autonomous Systems}
	\institution{Southern University of Science and Technology}
	\city{Shenzhen}
	\country{China}
}

\def\showcomments{}

\ifthenelse{\isundefined{\showcomments}}{
	\newcommand{\editnote}[3]{}
	
}{
	\newcommand{\editnote}[3]{\xspace\colorbox{#1}{\sffamily \smaller \textcolor{white}{~\faCommenting{}~#2~}}\textcolor{#1}{~#3}\xspace}
	
}

\definecolor{mygreen}{HTML}{02818a}
\definecolor{darkorange}{HTML}{FE6741}

\definecolor{nord0}{HTML}{2E3440}
\definecolor{nord1}{HTML}{3B4252}
\definecolor{nord2}{HTML}{434C5E}
\definecolor{nord3}{HTML}{4C566A}
\definecolor{nord4}{HTML}{D8DEE9}
\definecolor{nord5}{HTML}{E5E9F0}
\definecolor{nord6}{HTML}{ECEFF4}
\definecolor{nord7}{HTML}{8FBCBB}
\definecolor{nord8}{HTML}{88C0D0}
\definecolor{nord9}{HTML}{81A1C1}
\definecolor{nord10}{HTML}{5E81AC}
\definecolor{nord11}{HTML}{BF616A}
\definecolor{nord12}{HTML}{D08770}
\definecolor{nord13}{HTML}{EBCB8B}
\definecolor{nord14}{HTML}{A3BE8C}
\definecolor{nord15}{HTML}{B48EAD}

\newcommand{\code}[1]{{\small\color{violet}\text{\ttfamily#1}}}

\definecolor{summarybg}{HTML}{E1EDFC}
\definecolor{findingbg}{HTML}{FFF8E6}

\newcounter{rq}
\newenvironment{summary}{%
	\begin{tcolorbox}[
			arc=2mm,
			boxrule=0.0pt,
			left=2pt,
			right=2pt,
			top=2pt,
			bottom=2pt,
			colback=summarybg,
			colframe=summarybg
		]
		\textbf{\faLightbulbO\ RQ\refstepcounter{rq}\therq{} in Summary:}
		}{\end{tcolorbox}}

\newcounter{implication}

\titleformat*{\paragraph}{\bfseries}

\newcommand{\github}{\textsc{GitHub}\@\xspace}
\newcommand{\eg}[1]{(e.g., #1)}
\newcommand{\ie}[1]{(i.e., #1)}
\newcommand{\etal}[0]{et al.}

\newcommand{\projects}{11}
\newcommand{\testcases}{4,652}
\newcommand{\sampledmethods}{284}

\newcommand{\java}{\textsc{Java}\@\xspace}
\newcommand{\android}{\textsc{Android}\@\xspace}
\newcommand{\junit}{\textsc{JUnit}\@\xspace}

\newcommand{\evosuite}{\textsc{EvoSuite}\@\xspace}
\newcommand{\randoop}{\textsc{Randoop}\@\xspace}
\newcommand{\mockito}{\textsc{Mockito}\@\xspace}
\newcommand{\easymock}{\textsc{EasyMock}\@\xspace}
\newcommand{\powermock}{\textsc{PowerMock}\@\xspace}
\newcommand{\jmock}{\textsc{jMock}\@\xspace}
\newcommand{\maven}{\textsc{Maven}\@\xspace}
\newcommand{\gradle}{\textsc{Gradle}\@\xspace}
\newcommand{\netty}{\textsc{Netty}\@\xspace}

\newcommand{\mocksniffer}{\textsc{MockSniffer}\@\xspace}
\newcommand{\stubcoder}{\textsc{StubCoder}\@\xspace}
\newcommand{\rick}{\textsc{Rick}\@\xspace}
\newcommand{\moka}{\textsc{MOKA}\@\xspace}
\newcommand{\automock}{\textsc{AutoMock}\@\xspace}

\newcommand{\artifactlink}{\url{https://doi.org/10.5281/zenodo.14695509}}

\begin{abstract}
    Mock assertions provide developers with a powerful means to validate program behaviors that are unobservable to test assertions.
Despite their significance, they are rarely considered by automated test generation techniques.
Effective generation of mock assertions requires understanding how they are used in practice.
Although previous studies highlighted the importance of mock assertions, none provide insight into their usages.
To bridge this gap, we conducted the first empirical study on mock assertions, examining their adoption, the characteristics of the verified method invocations, and their effectiveness in fault detection.
Our analysis of \testcases{} test cases from \projects{} popular \java{} projects reveals that mock assertions are mostly applied to validating specific kinds of method calls, such as those interacting with external resources and those reflecting whether a certain code path was traversed in systems under test.
Additionally, we find that mock assertions complement traditional test assertions by ensuring the desired side effects have been produced, validating control flow logic, and checking internal computation results.
Our findings contribute to a better understanding of mock assertion usages and provide a foundation for future related research such as automated test generation that support mock assertions.

\end{abstract}

\begin{document}
\maketitle

% \ifthenelse{\isundefined{\editmode}}{
% }{
%     \input{sections/0_authors_ref.tex}
% }

\section{Introduction}
\begin{figure}[t]
	\centering
	\includegraphics[width=\linewidth]{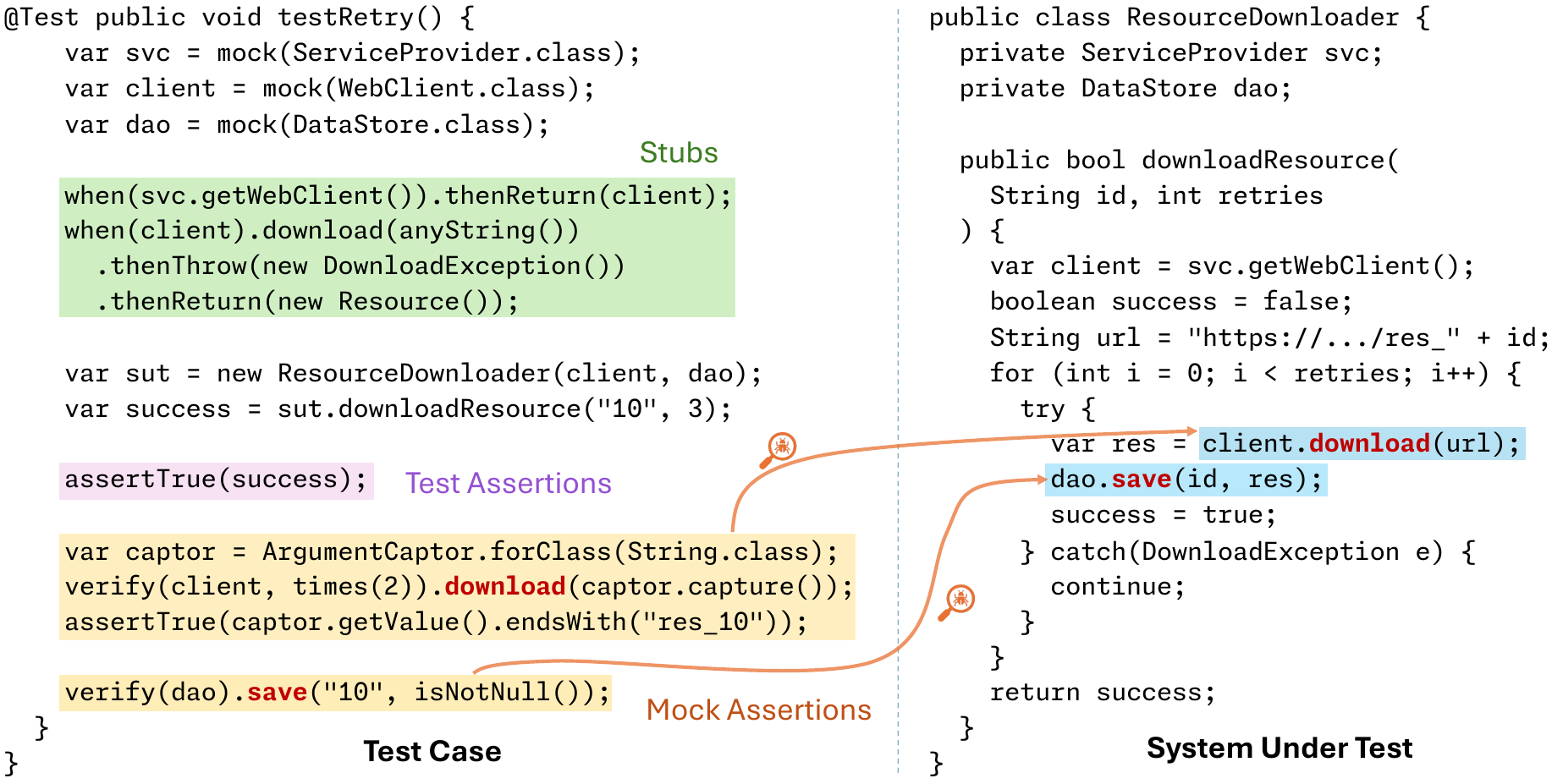}
	\caption{An Illustration of Mock Assertions and Test Assertions in Unit Testing.
		Mock assertions are leveraged to check if the URL used for downloading is correct and whether the resource has been saved to the database.
	}\label{fig:mock-assertion-illustration}
\end{figure}

Unit testing aims to verify the correctness of the system under test (SUT) in isolation.
However, in typical software, an SUT needs to interact with other components to fulfill its intended functionality.
When testing a SUT, developers often substitute its dependent components with test doubles~\cite{ACMDL:endo-testing}.
Test doubles are objects that simulate the behaviors of such dependent components in controlled ways.
An important role played by test doubles is mocking, where they record the method calls (along with their arguments) made to them, allowing developers to write \emph{mock assertions} to verify if the expected method invocations occurred during testing.
Mock assertions can validate program behaviors that are unobservable to \emph{test assertions}~\eg{those written using \junit{} or \textsc{AssertJ}}.
Consider the example in Figure~\ref{fig:mock-assertion-illustration}, where the SUT saves data in a database by calling a method of a database access object (DAO).
Since the data are passed to the method call to DAOs without changing any variables or return values that are accessible in the test case, the data value is unobservable to test assertions.
Therefore, test assertions cannot check whether the data written to the database is null.
Fortunately, developers can leverage mock assertions to verify the recorded method calls to a mock DAO.
By checking the arguments passed to the method call that sends out the data, developers can validate the correctness of database access.

Despite their importance in assuring software quality, mock assertions are rarely considered in existing automated test generation techniques.
The assertions constructed by test generators~\eg{\randoop{}~\cite{DBLP:conf/oopsla/PachecoE07}, \evosuite{}~\cite{DBLP:conf/qsic/FraserA11}} cannot predicate program behaviors that are unobservable to the test assertions and thus have weak test oracles~\cite{DBLP:conf/kbse/ShamshiriJRFMA15}.
Techniques like \textsc{AgitarOne}~\cite{Tool:agitarone} and \textsc{Rick}~\cite{DBLP:journals/corr/abs-2208-01321} do generate mock assertions.
However, they adopt a brute-force strategy to generate mock assertions for all method invocations of test doubles, resulting in aggressive mocking~\cite{DBLP:conf/kbse/ShamshiriJRFMA15} \ie{the generated tests contain excessive mock assertions that overfit the current implementation}, making the tests inflexible to changes in the SUT.
Shamshiri \etal{}~\cite{DBLP:conf/kbse/ShamshiriJRFMA15} reported that 31\% of the tests generated by \textsc{AgitarOne} raise false alarms due to aggressive mocking.
Indeed, during our study, we found such a brute-force strategy misaligned with developers' practice.
In fact, the aggressive mocking issue is an obstacle for \evosuite{} to adopt mock assertions~\cite{DBLP:conf/icst/ArcuriFJ17} as there is no effective mechanism to identify which interactions between the SUT and test doubles should be verified~\cite{DBLP:conf/icse/FazziniCCLKGO22}.

Generating effective mock assertions requires understanding their usage in practice.
Although several studies emphasize the importance of mock assertions~\cite{DBLP:conf/msr/SpadiniABB17, DBLP:journals/ese/SpadiniABB19,DBLP:conf/kbse/ZhuWWLCSZ20,DBLP:conf/icse/FazziniCCLKGO22,DBLP:journals/ese/XiaoZWLLWYW24}, none provides such insights.
To bridge this gap, we conducted the first empirical study on the usage of mock assertions.
Our study distills insights and empirical evidence for future research exploring the generation of appropriate mock assertions.
We also provide guidance for developers to write better mock assertions.
Specifically, we investigated:
\begin{enumerate*}
	\item the frequency of using mock assertions when using test doubles
	\item the characteristics of method invocations that are verified by mock assertions, and
	\item how mock assertions complement test assertions in fault detection
\end{enumerate*}.

In our study, we analyzed the usage of mock assertions in \testcases{} test cases from \projects{} large-scale, popular \java{} projects that use \mockito{}\footnote{\mockito{} is the most popular (used in over 80\% of the projects) mocking framework in \java{}~\cite{DBLP:journals/ese/SpadiniABB19,DBLP:conf/icse/FazziniCCLKGO22}}~\cite{Tool:mockito}.
We adopted open coding~\cite{lewis2015qualitative} to identify the common characteristics of the method invocations that are verified by mock assertions.
In addition, we performed mutation analysis~\cite{DBLP:journals/tse/JiaH11} to investigate how mock assertions complement test assertions in fault detection.
We had several interesting findings in our study.
For example, although mock assertions are used in 41\% of the test cases using test doubles, we found that verifying everything with mock assertions is not the state of the practice.
Instead, only 9\% of method invocations are verified (Section~\ref{sec:rq1}).
The contrast between the frequent adoption and the low verification ratio motivated us to investigate the characteristics of the method calls verified by mock assertions.
During our investigation, we identified three categories of methods whose invocations are commonly verified by developers (Section~\ref{sec:rq2}) and two types of common interactions between the SUT and the verified method invocations (Section~\ref{sec:rq3}).
Finally, we find that mock assertions complement test assertions by ensuring that the desired side effects have been produced, validating control flow logic, and checking internal computation results (Section~\ref{sec:rq4}).
This finding aligns with our identified characteristics of method invocations verified by mock assertions.
Currently, the identification of such characteristics still relies on manual efforts.
Future research can explore automated mechanisms to pinpoint critical method invocations during test execution and generate mock assertions to verify them accordingly.

In essence, we have made the following contributions in this paper:
\begin{itemize}
	\item To the best of our knowledge, we are the first to investigate the characteristics of mock assertion usage in practice.
	\item We identified three major categories of methods and two major categories of interactions that are often verified by mock assertions.
	      Our findings can shed light on the state of the practice of using mock assertions and provide guidance for both future researchers and developers.
	\item We investigated the fault detection capabilities of mock assertions and test assertions.
	      We found mock assertions complement test assertions by ensuring the desired side effects have been produced, validating control flow logic, and checking internal computation results.
	\item We constructed a dataset of method calls that are verified by developers with mock assertions.
	      We release it with our experimental data to facilitate future research endeavors.
	      Our research artifact is available at \artifactlink{}~\cite{artifact:zenodo/v1}.
\end{itemize}

\section{Background}\label{sec:background}

Test doubles and mock assertions are crucial components of software testing to deal with test dependencies.
In this section, we provide an overview of the roles played by test doubles and demonstrate the utilization of mock assertions through an illustrative example.

\subsection{Test Doubles in Unit Testing}

In software testing, test doubles~\cite{ACMDL:endo-testing,DBLP:conf/icse/FazziniCCLKGO22} are simulated objects that represent the actual dependencies during testing.
They can be designed to mimic the behavior of real objects in controlled environments or utilized to verify interactions between various system components without directly involving the actual dependencies.

Test doubles, often created using mocking frameworks like \mockito{} in \java{}, can be configured to return specific values, simulate error conditions, or verify that certain methods are called with the expected arguments.
Depending on their usage, test doubles can play the following roles.
\begin{itemize}
	\item \textbf{Mock:}
	      Mock objects are configured with expectations about the interactions they will have with the SUT.
	      They record the method calls made to them and enable developers to verify if the anticipated interactions have occurred.
	      Also, the arguments used in these methods calls are also recorded for further checking.
	      For example, the test double \code{dao} in Figure~\ref{fig:mock-assertion-illustration} is a mock.
	\item \textbf{Spy:}
	      Spy objects are a specialized type of mock object that wraps a real object.
	      In addition to recording method calls, they also call the real methods in the wrapped objects.
	\item \textbf{Stub:}
	      Stub objects provide predefined responses to method calls.
	      They return predefined values and are used to simulate specific behaviors of dependencies.
	      For example, the test double \code{svc} in Figure~\ref{fig:mock-assertion-illustration} is a stub.
	\item \textbf{Fake:}
	      Fake objects are more sophisticated versions of stubs.
	      Instead of returning predefined values, fakes offer a lightweight implementation of the actual component solely for testing purposes, such as simulating a database by storing data in an array.
	\item \textbf{Dummy:}
	      Dummies are objects with specific types or implementing certain interfaces.
	      They lack any logic and are merely used to fulfill method signatures during testing.
	      Neither the test nor the SUT interacts with them.
\end{itemize}

In practice, test doubles created with common frameworks like \mockito{} can play one or more of these roles.
For example, in Figure~\ref{fig:mock-assertion-illustration}, the test double \code{client} is used as both stub and mock.
In this paper, we focus on mock assertions, which are used with the test doubles of the role mock or spy.

\subsection{Mock Assertions vs.
	Test Assertions}\label{sec:ma-example}

In unit testing using test doubles, there are two types of assertions that can be used to predicate on the program behaviors to assure the correctness of SUT.
\begin{itemize}
	\item \textbf{Test Assertions.}
	      Test assertions are executable boolean expressions in the test that predicate the values of variables.
	      The variables predicated by test assertions are either the output returned from the SUT or the states mutated by the SUT.
	      In practice, developers use common testing frameworks such as \junit{} and \textsc{AssertJ} to write test assertions.
	\item \textbf{Mock Assertions.}
	      Mock assertions are assertions that predicate on the recorded method invocations on mock or spy objects.
	      They enable developers to check if the interactions between the SUT and test doubles have occurred in expected ways.
	      Specifically, developers can check whether a certain method of a test double has been invoked, the number of invocations, and whether the arguments used for the method invocations match the expected values.
	      In addition, mock assertions allow developers to capture the value of arguments and further predicate them with \junit{} assertions.
\end{itemize}
Mock assertions complement test assertions by checking the program behaviors that cannot be observed by test assertions.
Figure~\ref{fig:mock-assertion-illustration} shows an example of unit testing with mock assertions using the \mockito{} framework.
The SUT \code{ResourceDownloader} downloads a resource using a \code{WebClient} and stores the result in the database with \code{DataStore}.
Since the download may fail, callers of the method \code{downloadResource} can specify the number of maximum retries via parameter \code{retries}.
The test case \code{testRetry} validates the retry logic.
Before calling the method \code{downloadResource}, a mock \code{WebClient} is configured to simulate the scenario where the method \code{download} will fail for the first invocation and succeed in the second try.
In this case, in addition to returning \code{true} to indicate a successful operation, the method \code{downloadResource} should try to download exactly twice and finally store the resource using \code{DataStore}.
Since the download operation should be successful given such a scenario, developers checked the return value of \code{downloadResource} with a test assertion.

However, this test assertion alone cannot ensure the correctness of the implementation as several key operations are unchecked.
First, the resource should be downloaded exactly twice since the first download failed and the second download was successful.
Second, the correct URL should be used for the download.
Third, the downloaded resource should be saved into the database.
These operations are unobservable in the test since the relevant data is not returned by \code{downloadResources} or accessible via any getters.
Therefore, they cannot be checked by test assertions.
Fortunately, these operations manifest as the interactions between the SUT and the test dependencies.
In this example, the download and database access are done by calling the methods in \code{WebClient} and \code{DataStore}.
Therefore, developers use mock assertions to ensure these expected behaviors have occurred.
To validate the retry logic, developers use a mock assertion to check if \code{client.download()} was called twice.
The argument passed to the method \code{download} is also captured with an \code{ArgumentCaptor}, and the captured value is further checked with a \junit{} assertion.
Although such an assert statement is similar to test assertions, it is considered as part of the mock assertions since it predicates the value captured by a mocked method call.
In addition, to ensure the resource is stored in the database, developers use another mock assertion to check if \code{dao.save()} was invoked as expected.

In general, mock assertions enable developers to predicate on the behaviors that are not directly observable from the test.
This is done by checking the recorded method calls (and their arguments) on mock objects.
Mock assertions complement test assertions and make the test stronger.

\section{Empirical Study Design}\label{sec:study-design}
In this section, we outline the design of our empirical study design.
Specifically, we outline the research questions and the rationale behind them, and we present our data collection process.

\subsection{Research Questions}
The goal of this paper is to guide developers in effectively using mock assertions and offer insights to future researchers for developing automated techniques to generate and maintain mock assertions.
To achieve our goal, we investigated the following four research questions (RQs):
\begin{itemize}
	\item \textbf{RQ1: Adoption.} \textit{How frequently do developers use mock assertions when using test doubles?}
	      This RQ focuses on understanding the frequency of mock assertion adoption among developers, providing insights into the prevalence of this assertion type in practice.
	\item \textbf{RQ2: Method Types.} \textit{What kinds of method calls are verified by mock assertions?}
	      This RQ explores the types of methods that developers typically verify their invocations using mock assertions, offering understandings of the specific areas in which mock assertions are used.
	\item \textbf{RQ3: Interactions.} \textit{How do the SUTs interact with the methods verified by mock assertions?}
	      In practice, developers use mock assertions to verify interactions between the SUT and the test dependencies~\cite{DBLP:journals/ese/SpadiniABB19,DBLP:conf/kbse/ZhuWWLCSZ20,DBLP:conf/icse/FazziniCCLKGO22}.
	      Therefore, we seek to investigate the characteristics of these interactions, aiming to gain insights into the interactions that are of particular interest to developers when using mock assertions.
	\item \textbf{RQ4: Fault Detection.} \textit{How do mock assertions complement test assertions in fault detection?}
	      This research question examines the role of mock assertions in detecting potential faults.
	      We specifically focus on test cases where both assertion types are employed.
	      Our objective is to shed light on the areas where mock assertions can complement test assertions, thereby contributing to the assurance of software reliability.
\end{itemize}

\subsection{Data Collection}
To support the investigation of the four research questions, we constructed a dataset of unit tests that interact with test doubles.
We detail our data collection process as follows.

\paragraph{Project Selection.}
We used the query \textit{language:java archive:false pushed:>2024-01-01} to search for repositories on \github{} to identify actively maintained Java projects that have been updated since 2024.
The search was done in March 2024 and it returned over 2 million projects.
We sorted the search results by the number of stars, which is an indicator of popularity.
Popular projects are widely used and are more likely to attract experienced contributors following best practices~\eg{appropriate use of mock assertions}.
Prioritizing popular projects for our study allows us to generate better findings for developers and future researchers.
Beyond popularity, we applied the following criteria to ensure subject quality:
\begin{enumerate}
	\item The project contains at least 10k lines of \java{} code, filtering out smaller toy projects.
	\item We manually reviewed the README file to make sure that the project is not a tutorial or an Android project, as such projects fall outside the scope of this study.
	\item It declares \mockito{} as a dependency in the \maven{}
	      POM file or \gradle{} build scripts because our data extraction tool is implemented based on \mockito{}, the most widely used mocking framework in \java{}~\cite{DBLP:conf/qsic/MostafaW14,DBLP:conf/icse/FazziniCCLKGO22}.
\end{enumerate}
We cloned the repositories of the most stared 20 projects meeting the above criteria.
We followed their documentation to build their latest release version and run the tests.
We discarded a project if we were unable to build it or run the test.
Finally, we got a list of \projects{} projects, as detailed in Table~\ref{tab:subjects}.
These projects are large-scale, sophisticated, and span major application domains of the \java{} programming languages \eg{bug data, web, database}, which bolsters our findings in generalizing to these domains.

\paragraph{Unit Tests Identification.}
In this paper, we focus on unit tests, since test doubles are used primarily in unit tests to isolate the SUT from its dependencies~\cite{DBLP:journals/ese/SpadiniABB19}.
To identify unit tests in these projects, we adopted an approach similar to that used in recent studies~\cite{DBLP:conf/kbse/ZhuWWLCSZ20,DBLP:journals/ese/XiaoZWLLWYW24}.
Specifically, we identify the methods annotated with the \junit{} \code{@Test} annotation as test cases.
To ensure that a test is indeed a unit test, we follow the pattern used by the \junit{} test runner.
We infer the name of the SUT by stripping the prefix or suffix \code{Test} from the name of the test class and check if there exists such a class in the same package~\cite{DBLP:conf/kbse/ZhuWWLCSZ20,arxiv:tufano2020unit}.
Then we ran the tests with an instrumented \mockito{} to check if they interact with the test doubles~\ie{invoke at least one method of test doubles}.
We aim to exclude test cases that use test doubles as dummies since they are out of the scope of this paper.
As shown in Table~\ref{tab:subjects}, we identified \testcases{} test cases that interact with test doubles.

\paragraph{Method Call Extraction.}
To construct a dataset of method calls that are verified by developers, we ran all the identified test cases for data extraction.
We opted for a dynamic analysis approach due to its ability to capture runtime interactions between the SUT and mock objects during test execution.
Dynamic analysis allows us to observe actual method calls and their contexts, which enhances the accuracy of our findings regarding mock assertion usage.
Similar approaches have been adopted by recent studies~\cite{DBLP:conf/kbse/ZhuWWLCSZ20,DBLP:conf/icse/FazziniCCLKGO22} for data collection.

During test execution, we used an instrumented \mockito{} to collect the runtime interactions between the SUT and test doubles.
Specifically, we attached a debugger to the test runner and set a breakpoint in \code{MockitoCore}, which is used by \mockito{} internally to generate test doubles.
When the breakpoint is hit, we mutated the \code{MockSettings} object to inject a \code{InvocationListener} to track method calls to the test double being created.
After the test execution, we used the API \code{Mockito.mockingDetails} to inspect the interactions between the SUT and the test doubles, and identify the method calls verified by mock assertions.
We disabled garbage collection for the test doubles to enable after-test inspection.

The information collected during test execution is gathered to form our dataset.
Each entry of our dataset corresponds to a method invocation and it contains the following information:
\begin{enumerate*}
	\item the type and object ID of the test double,
	\item the signature of the method call,
	\item the stack trace of the method call,
	\item a label indicating if developers stubbed a custom return value to the method call, and
	\item a label indicating if the method call is verified
\end{enumerate*}.
We extracted stack traces at the injected breakpoints and retrieved stub/verify labels from \mockito{} API \code{MockingDetails}.
We eliminated duplicates from the dataset by removing the method calls in the same test case with identical method signatures, mock object IDs, and stack traces, aiming to prevent skewing the dataset by repeated method calls, such as those within loops.

Table~\ref{tab:mock-assertions} presents statistical information of the method invocations to test doubles.
During test execution, the SUTs frequently interact with the mock objects.
The 4,652 test cases made 40,639 method invocations to the mock objects.
Such a frequency aligns with a recent study~\cite{DBLP:conf/icse/FazziniCCLKGO22} conducted on \android{} applications.
Among these method invocations, developers specified a custom return value for 24,928 (61\%) of them.
In addition, developers verified 3,621 of them with mock assertions.
Our subsequent analysis will be based on these verified method invocations.

\begin{table}[t]
	\centering
	\caption{Overview of Studied Projects and Test Cases}\label{tab:subjects}
	\smaller
\begin{tabularx}{\linewidth}{lLr|rrr|r}
\toprule
\multirow{3}{*}{\textbf{Project}} & 
\multirow{3}{*}{\textbf{Application Domain}} &
\multirow{3}{*}{\textbf{LoC in \java{}}} &
\multicolumn{3}{c|}{\textbf{Unit Test Cases}} &
\multirow{3}{*}{\textbf{Test Doubles Created}}\\
\cmidrule(l{1pt}r{1pt}){4-6}
&&&
\textbf{Total} &
\textbf{w/ TD}\textsuperscript{1} &
\textbf{w/ MA}\textsuperscript{2} & 
\\
\midrule
\textsc{Camel}          & Application integration & 1,874k  & 12,535 & 416   & 180   & 1,159  \\
\textsc{CXF}            & Web services            & 837k    & 5,592  & 446   & 94    & 1,328  \\
\textsc{Dubbo}          & RPC framework           & 286k    & 3,481  & 207   & 52    & 404    \\
\textsc{Hadoop}         & Big data                & 2,726k  & 14,013 & 995   & 375   & 3,221  \\
\textsc{Hazelcast}      & In-memory data grid     & 1,414k  & 22,482 & 185   & 64    & 389    \\
\textsc{Kafka}          & Event streaming         & 190k    & 7,644  & 575   & 241   & 1,188  \\
\textsc{MyBatis3}       & Database ORM            & 64k     & 1,664  & 301   & 185   & 365    \\
\textsc{Neo4j}          & Graph database          & 883k    & 5,457  & 777   & 435   & 2,179  \\
\textsc{Ozone}          & Object storage          & 1,165k  & 3,031  & 165   & 67    & 482    \\
\textsc{Spring Boot}    & Web framework           & 428k    & 4,568  & 512   & 185   & 920    \\
\textsc{Storm}          & Real-time processing    & 337k    & 594    & 73    & 39    & 263    \\
\midrule
\textbf{Total}          &                         & 10,204k & 81,061 & 4,652 & 1,917 & 11,898 \\
\bottomrule
\end{tabularx}
\begin{tablenotes}
    \scriptsize
    \item 1: With Test Doubles -- Test cases invoking at least one method of a test double.
    \item 2: With Mock Assertions -- Test cases containing at least one mock assertion.
\end{tablenotes}

\end{table}

\section{RQ1: How Frequently Do Developers Use Mock Assertions When Using Test Doubles?}\label{sec:rq1}

Table~\ref{tab:subjects} presents the demographics of our dataset.
Among the 4,652 test cases that call at least one method of mock objects, 1,917 contain at least one mock assertion, representing 41\% of the total.
This observation underscores the widespread adoption of mock assertions by developers when using mock objects, emphasizing their significance in ensuring the correctness of software.

\begin{table}[t]
	\centering
	\caption{Overview of Method Invocations on Test Doubles}\label{tab:mock-assertions}
	\smaller
\setlength{\tabcolsep}{3pt}
\begin{tabularx}{\linewidth}{lr|RRR|RRR}
\toprule
\multirow{3}{*}{\textbf{Project}} &
\multirow{3}{*}{\textbf{TDs Created}} &
\multicolumn{3}{c|}{\textbf{Unique Invocations\textsuperscript{1}}} &
\multicolumn{3}{c}{\textbf{Unique Methods\textsuperscript{2}}}
\\
\cmidrule(l{1pt}r{1pt}){3-5}
\cmidrule(l{1pt}r{1pt}){6-8}
&&
\textbf{Total} & \textbf{Stubbed} & \textbf{Verified} &
\textbf{Total} & \textbf{Stubbed} & \textbf{Verified}
\\
\midrule
\textsc{Camel}         & 1,159  & 2,966  & 1,982 (67\%)  & 467 (16\%)  & 470   & 274   & 126   \\
\textsc{CXF}           & 1,328  & 4,323  & 3,191 (74\%)  & 209 (5\%)   & 511   & 344   & 74    \\
\textsc{Dubbo}         & 404    & 1,934  & 1,266 (65\%)  & 138 (7\%)   & 313   & 161   & 66    \\
\textsc{Hadoop}        & 3,221  & 16,612 & 9,289 (56\%)  & 879 (6\%)   & 1,268 & 599   & 194   \\
\textsc{Hazelcast}     & 389    & 826    & 675   (82\%)  & 74  (9\%)   & 116   & 89    & 22    \\
\textsc{Kafka}         & 1,188  & 5,513  & 3,196 (58\%)  & 463 (8\%)   & 537   & 309   & 134   \\
\textsc{MyBatis3}      & 365    & 578    & 403   (70\%)  & 76  (14\%)  & 162   & 104   & 33    \\
\textsc{Neo4j}         & 2,179  & 4,457  & 2,554 (57\%)  & 917 (21\%)  & 770   & 385   & 290   \\
\textsc{Ozone}         & 482    & 1,241  & 929   (75\%)  & 94  (8\%)   & 151   & 102   & 40    \\
\textsc{Spring Boot}   & 920    & 1,665  & 1,168 (70\%)  & 221 (14\%)  & 337   & 220   & 101   \\
\textsc{Storm}         & 263    & 524    & 275   (52\%)  & 83  (16\%)  & 109   & 53    & 30    \\
\midrule
\textbf{Total}         & 11,898 & 40,639 & 24,928 (61\%) & 3,621 (9\%) & 4,628 & 2,576 & 1,080 \\
\bottomrule
\end{tabularx}
\begin{tablenotes}
    \scriptsize
    \item 1: Uniquely identified by test case, method signature, object ID, and stack trace.
    \item 2: Uniquely identified by method signature and mock object type.
\end{tablenotes}

	\vspace{-1em}
\end{table}

Notably, despite the prevalent usage of mock assertions and the frequent interaction between the SUTs and mock objects, we observed a relatively low verification rate for method calls.
As indicated in the fifth column of Table~\ref{tab:mock-assertions}, the verification ratio varies from 5\% to 21\% across our subjects, with \textsc{Neo4j} being the most prevalent and \textsc{CXF} being the least.
On average, only 9\% of method calls made to mock objects are verified by mock assertions.
The result shows that verifying every method calls on test doubles with mock assertions is not the state of the practice.
The disparity between the prevalent usage and the low verification ratio prompted us to conduct a deeper analysis to elucidate the specific usage patterns of mock assertions.

Interestingly, we found that only 1,032 (4\%) of the stubbed invocations are verified, while 2,589 (17\%) of the non-stubbed invocations are verified.
A chi-square test~\cite{Pearson1992} of independence was performed to examine the relation between the usage of stub and mock.
The relation between these two variables was significant (\(p < 0.05\)).
Within the same test double, developers are unlikely to verify the calls to the methods used for the role stub.
While a test double can assume various roles, it is rare for these roles to be fulfilled by the same method.

\begin{summary}
	Mock assertions are used in 41\% of test cases using test doubles.
	However, only 9\% (on average) of the method invocations on test doubles were verified by mock assertions.
	The result shows that verifying every method calls on test doubles with mock assertions is not the state of the practice.
	In addition, within the same test double, it is uncommon for a method to fulfill multiple roles.
\end{summary}

\section{RQ2: What Kinds of Method Calls Are Verified by Mock Assertions?}\label{sec:rq2}
RQ2 aims to understand the types of methods developers verify using mock assertions.
This section details our approach to identifying the characteristics of these methods and presents our findings.

\subsection{Empirical Study Approach}
As illustrated in Table~\ref{tab:mock-assertions}, a total of 3,621 unique method invocations are verified by mock assertions, which correspond to 1,080 distinct methods.
To gain deeper insights into the characteristics of verified methods \ie{whose invocations are verified by mock assertions}, we randomly sampled a statistically significant subset of 284 methods, ensuring a confidence level of 95\% and a margin of error of 5\%.
To understand the difference between the verified methods and not verified methods, we further sampled 347 methods from the remaining 3,548 (4,628 - 1,080) for analysis.
Such a sample size ensures a confidence level of 95\% and a margin of error of 5\%.

We employed an open coding approach~\cite{lewis2015qualitative} to systematically categorize these methods.
Initially, we created tags based on JavaDoc, implementations, and comments associated with each selected method.
These tags describe the roles and behaviors of the methods.
Following this tagging process, we performed two rounds of axial coding.
Specifically, we grouped method with similar tags together and came up with a higher-level tag to describe them.
For example, we tagged \code{FileChannel.truncate()} with \textit{File System} during open coding.
Later in axial coding, it was merged into \textit{I/O} and finally merged into \textit{External Resource}.
Two authors participated in the task and they discussed the results in their meetings.
On disagreement, a third author joined to resolve the conflict.
We finally got five categories from the coding process.
These categories reflect the behaviors and roles of the methods whose invocations are verified (or not) by developers with mock assertions.
After the coding process, we performed Chi-square tests of independence to examine whether the methods in each category are (un)likely to be verified.

\subsection{Empirical Study Results}\label{sec:rq2-results}

\begin{table}[t]
	\centering
	\caption{Categorization of Methods Types Verified (or Not) by Mock Assertions.
		Chi-square tests of independence examined the relation between each category's membership and the method's verification.
	}\label{tab:method-types}
	\smaller
\setlength{\tabcolsep}{3pt}
\begin{tabularx}{\linewidth}{lrrrl}
	\toprule
	\textbf{Category\textsuperscript{1}} &
	\textbf{Verified}                    &
	\textbf{Not Verified}                &
	\(\chi^2 (p < 0.05)\)                &
	\textbf{Description}                                                                                                                                 \\
	\midrule
	External Resource                    & 130 (46\%) & 30 (9\%)   & 113.75                             & Interact with I/O, concurrency, database, etc. \\
	State Mutator                        & 79 (28\%)  & 47 (14\%)  & 19.91                              & Changes the internal fields of the object.     \\
	Callback                             & 39 (14\%)  & 2 (0.06\%) & 44.49                              & Event handlers, listeners, etc.                \\
	Accessor                             & 41 (14\%)  & 261 (75\%) & 231.10                             & Returns the internal state of the object.      \\
	Others                               & 7 (3\%)    & 10 (3\%)   & 0.1036                             & Do not fall into any of the categories above.  \\
	\midrule
	\textbf{Sample Size}                 & 284        & 347        & Significant when \(\chi^2 > 3.84\) & CL=95\%, ME=5\%                                \\
	\bottomrule
\end{tabularx}

\begin{tablenotes}
	\item \scriptsize * These categories may overlap.
	For example, a callback may also be a state mutator.
\end{tablenotes}

\end{table}

Table~\ref{tab:method-types} shows the categorization of the method types we have identified.
In total, we identified four categories of the methods.
We present each of these categories in detail.

\paragraph{External Resources (130/\sampledmethods{}).}
This category encompasses methods that interact with external resources, which are crucial for enabling applications to communicate with the outside world.
Specifically, it includes the methods that
\begin{enumerate*}
	\item manage file systems, transmit data over networks, generate logs, access streams,
	\item interact with resources provided by operating systems \eg{processes, threads, authentication, containers, and hardware}, and
	\item access external databases, manages database connections, executing SQL queries, and modifying data in key-value stores.
\end{enumerate*}

\begin{figure}[t]
	\centering
	\includegraphics[width=\linewidth]{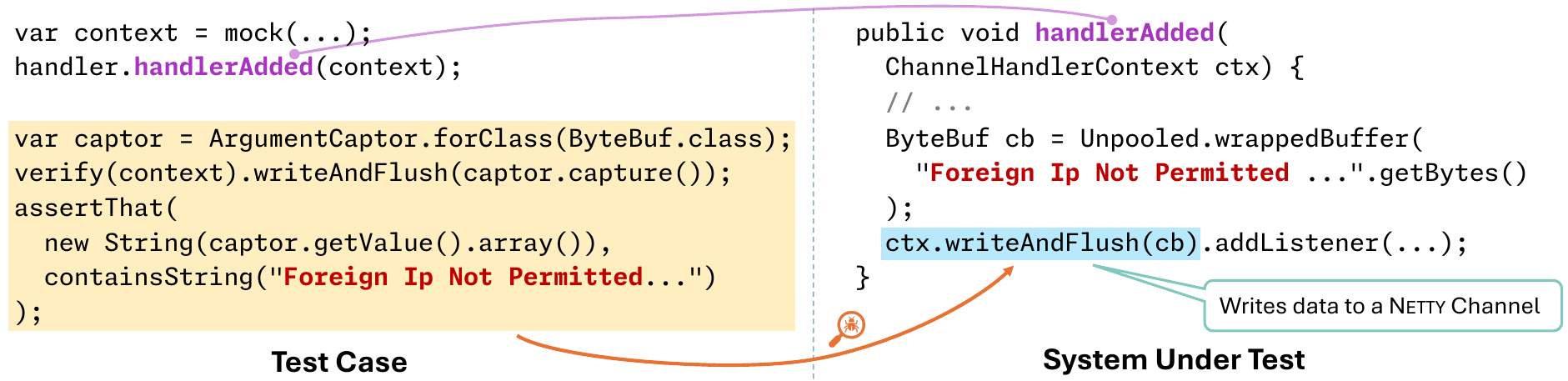}
	\caption{A Mock Assertion Verifying an Invocation to a Method that Writes Data to a \textsc{Netty}
		Channel (Project \textsc{Dubbo}).
		Developers check the argument to ensure the correctness of the data being written.
	}
	\label{fig:method-env-example}
\end{figure}

Developers use mock assertions to verify that methods interacting with external resources are called with the appropriate parameters such that anticipated interactions with the environment take place.
For instance, Figure~\ref{fig:method-env-example} illustrates an example from project \textsc{Dubbo}, where developers leverage mock assertions to check the correctness of data written to a \netty{} channel.
In this example, the test case invokes the method \code{handlerAdded} in SUT with a mock \code{ChannelHandlerContext}.
Inside the SUT, the error message is encoded into a buffer and subsequently written to the mock \netty{} channel by calling its method \code{writeAndFlush}.
Back in the test case, developers first utilize a mock assertion to ensure that the method \code{writeAndFlush} is invoked.
Then, they employ an \code{ArgumentCaptor} to extract the argument and verify that it contains the expected error message using a \junit{} assertion.
Such a mock assertion ensures the correct data has been transmitted over the network.

Among the 284 randomly sampled methods, 130 (46\%) interact with external resources.
In comparison, methods in this category only count for 9\% of the methods that are not verified.
The result of the Chi-square test was significant, which shows that methods interacting with external resources are more commonly verified.
By checking the method calls to these methods with mock assertions, developers ensure the soundness of the integration points between the SUT and the external environment.
Ultimately, this practice enhances test reliability and fosters confidence in the system’s interactions with external dependencies.

\paragraph{State Mutators (79/\sampledmethods{}).}
This category includes methods that mutate the state of the test dependencies, which is essential for program state transitions within the system.
Such methods include
\begin{enumerate*}
	\item setters that update the value of fields, and
	\item domain-specific mutators that change the object's internal state.
\end{enumerate*}

\begin{figure}[t]
	\centering
	\includegraphics[width=\linewidth]{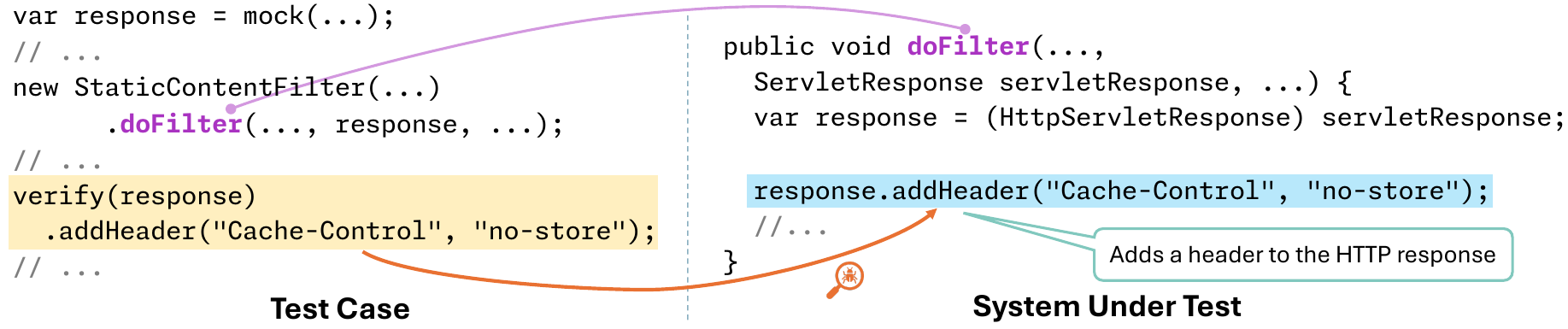}
	\caption{A Mock Assertion Verifying an Invocation to a State Mutator that Inserts a Header to an HTTP Response (Project \textsc{Neo4j}).
		Developers check the correctness of the arguments.
	}
	\label{fig:method-mutator-example}
\end{figure}

Verifying method calls to state mutators is essential for ensuring that the correct state transitions have taken place.
For example, Figure~\ref{fig:method-mutator-example} shows an example from the \textsc{Neo4j} project.
In the test case, the developer calls the method \code{doFilter} in SUT with a mock \code{response}.
Inside the SUT, an HTTP header \code{"Cache-Control: no-store"} is added to the response via the method \code{addHeader}, which mutates a private field of \code{ServletResponse} by inserting the given header value.
Afterwards, the developer uses a mock assertion to ensures that the method has been invoked with the expected key and value.
Such a mock assertion ensure the correct header has been added.

Among the 284 randomly sampled methods, 79 (28\%) are classified as mutator methods.
In comparison, methods in this category only count for 14\% of the methods that are not verified.
The result of the Chi-square test was significant, which shows that state mutators are more commonly verified.
By employing mock assertions in conjunction with these methods, developers can effectively test the correctness of state transitions within their applications.
This practice reinforces the reliability of the system by ensuring that state changes do not lead to unintended side effects.

\paragraph{Callbacks (39/\sampledmethods{}).}
This category includes callback methods that function as event handlers or listeners, enabling applications to respond to events.
Callback methods are invoked in response to user actions, system events, or the completion of asynchronous tasks.
These methods may or may not interact with external resources or mutate the state of the test dependency, depending on the actual implementation.

\begin{figure}[t]
	\centering
	\includegraphics[width=\linewidth]{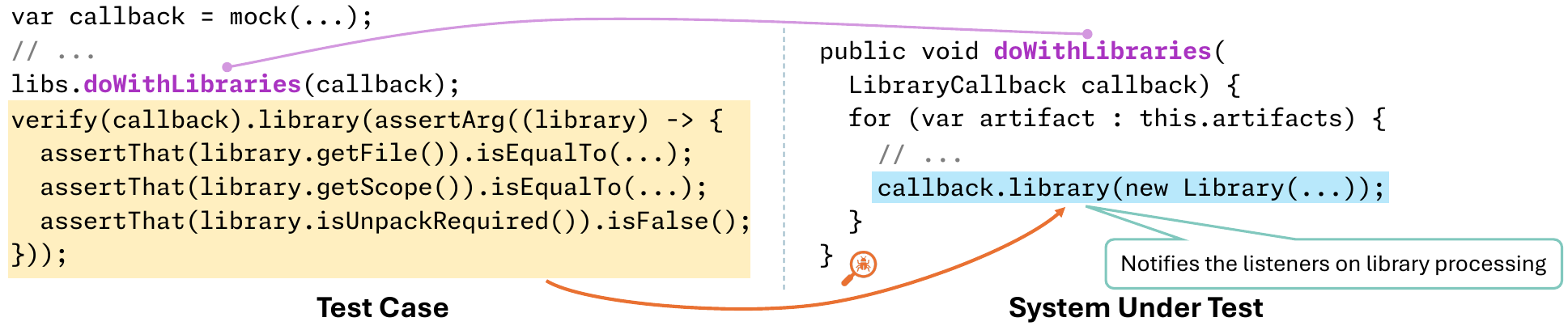}
	\caption{A Mock Assertion Verifying an Invocation of a Callback Method Triggered During Library Processing (Project \textsc{SpringBoot}).
		Developers further check if the \code{library} passed to the callback is correct.
	}
	\label{fig:method-callback-example}
\end{figure}

Developers verify invocations to callback methods to ensure that the event-handling pipeline is correctly established.
Figure~\ref{fig:method-callback-example} illustrates an example from the \textsc{SpringBoot} project.
In the test case, the developer invokes the method \code{doWithLibraries} in SUT with a mock \code{callback} to process library artifacts.
Inside the SUT, the method \code{library} is called when each artifact has been processed to notify the components that are interested in library processing.
In the test case, developers use a mock assertion to confirm that the callback method has been invoked.
They also extract the argument and further validate it with three \textsc{AssertJ} assertions.

Among the 284 randomly sampled methods, 39 (14\%) are classified as callback methods.
In comparison, methods in this category only count for 0.06\% of the methods that are not verified.
The result of the Chi-square test was significant, which shows that callback methods are more commonly verified.
Although this ratio may seem low, these methods are crucial for establishing the event-handling mechanism within applications.
By validating the invocation of callback methods, developers can ensure that the SUT correctly fires events to interested components.

\paragraph{Accessors (41/\sampledmethods{}).}
This category includes methods that are used to retrieve an external state via the test dependency.
Methods in this category include
\begin{enumerate*}
	\item methods that return a value based on the argument as a query key \eg{\code{getHeader}}, and
	\item getters that solely return the value of a field.
\end{enumerate*}

Among the 284 randomly sampled methods, 41 (14\%) fall under the accessor category.
In comparison, methods in this category count for 75\% of the methods that are not verified.
The result of the Chi-square test was significant, which shows that accessor methods are not commonly verified.
Although verifying method calls to accessors is not common, sometimes they can serve as a proxy for observing the actual execution path in the SUT and are important for revealing control flow discrepancies.
Such cases will be discussed in Section~\ref{sec:rq3} and Section~\ref{sec:rq4}.

\begin{summary}
	Developers use mock assertions to verify methods in three categories: method interacting with external resources (46\%), validating interactions with external systems; state mutators (28\%), validating program state transitions; and callbacks (14\%), ensuring event handlers are triggered properly.
	In comparison, accessors are unlikely to be verified.
\end{summary}

\section{RQ3: How Do the SUTs Interact With the Methods Verified by Mock Assertions?}\label{sec:rq3}
In unit testing, the SUT frequently interacts with the test doubles~\cite{DBLP:conf/icse/FazziniCCLKGO22} and developers use mock assertions to verify them~\cite{DBLP:conf/kbse/ZhuWWLCSZ20,DBLP:journals/ese/SpadiniABB19}.
Therefore, RQ3 aims to identify the characteristics of the interactions between the SUT and the verified method calls.

\subsection{Empirical Study Approach}
To answer RQ3, we randomly sampled a subset of 348 method calls from the 3,621 unique method calls as shown in Table~\ref{tab:mock-assertions}, ensuring a confidence level of 95\% and a margin of error of 5\%.

Similarly as in RQ2, we adopted the open coding approach~\cite{lewis2015qualitative} to identify the common characteristics of the method invocations made to the test doubles.
In RQ3, our focus shifts from the behavior of the methods themselves to the interactions between the SUT and the method invocations.
To systematically document these interactions, we characterized each method call based on its specific role in the context of the SUT's execution.
This involved analyzing the data flow and control flow patterns with each method call, allowing us to capture how information flow and control flow logic in the methods affect the method invocation within the SUT.
Based on the observed behaviors and roles of the method calls, we applied descriptive tags to categorize them.
Following this tagging process, we analyzed the tags to identify and merge similar categories, ultimately consolidating them into two major categories: one related to data flow and the other to control flow.
These categories reflect the dual nature of the interactions, highlighting the significance of both the information transfer and the decision-making processes that influence the method calls.

\subsection{Empirical Study Results} \label{sec:rq3-results}

Among the 348 sampled method calls, we identified two primary categories of interactions between the SUT and the mocked method call.
They are related to control flow and data flow, respectively.

\paragraph{Conditional Invocation (181/348).}
The first category of interactions we observe relates to the control flow characteristics of method calls.
These calls are executed conditionally, depending on the test input or the internal states of the SUT.
Examples of such conditional invocations include:
\begin{enumerate*}
	\item governed by \code{if}, \code{switch}, or loop constructs,
	\item occurring within exception handlers, and
	\item preceded by an early \code{return} or \code{throw}.
\end{enumerate*}

\begin{figure}[t]
	\centering
	\includegraphics[width=\linewidth]{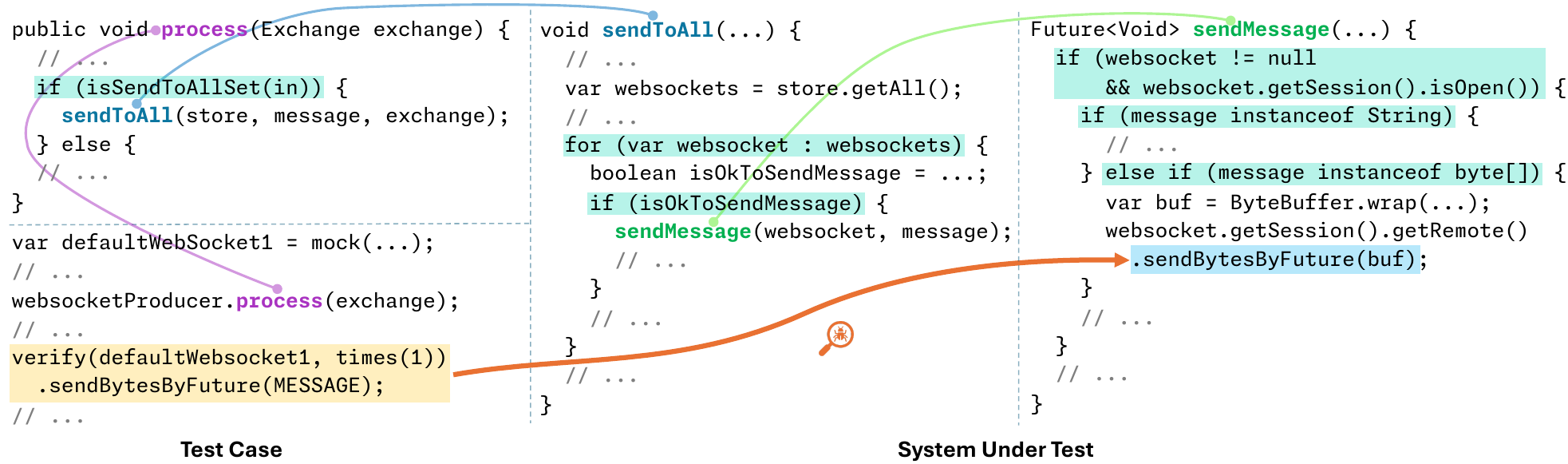}
	\caption{A Mock Assertion Verifying a Method Call that is Made Conditionally (Project \textsc{Camel}). The method invocation is guarded by multiple branch conditions and loops.}
	\label{fig:controlflow-example}
	% \vspace{-1em}
\end{figure}

Figure~\ref{fig:controlflow-example} illustrates an example from the \textsc{Camel} project.
In this test case, the developer calls the \code{process} method of the SUT.
After executing method calls to \code{sendToAll} and \code{sendMessage}, the execution ultimately dispatches a call to \code{sendBytesByFuture} of the object \code{defaultWebSocket1}.
This method call resides deep within nested branch conditions, governed by four \code{if} statements and a \code{for} loop.
The developer employs a mock assertion to verify whether this method was invoked, ensuring the correctness of the \code{if} statements and the \code{for} loop.
By using the presence of this method call as a proxy, developers can ascertain that the branch conditions (as highlighted in Figure~\ref{fig:controlflow-example}) are correctly implemented, thereby enhancing the reliability of the SUT.

Among the 348 sampled method calls, 181 (52\%) fall into this category.
The frequencies of these conditional invocations serve as a proxy to the actual execution path.
For example, a method call guarded an \code{if} statement is executed only when the conditions of the \code{if} statement is met.
Verifying such method invocations with mock assertion can help ensure the correctness of branch conditions.
As mentioned by a \textsc{Hadoop} developer in Pull Request \#1146\footnote{\textsc{Hadoop} HDDS-1366: \url{https://github.com/apache/hadoop/pull/1146\#discussion\_r312278997}}, \textit{Method call verification is generally used for mocked methods so that we know the code path went through that}.
In general, developers can validate the correctness of the implemented control flow logic by verifying the method calls that are made conditionally.

\paragraph{Data Consumers (179/348).}
The second category of interactions we observed pertains to the data flow aspect of method calls.
Specifically, two characteristics are found for these method calls:
\begin{enumerate*}
	\item in the SUT, internal computation results are passed as arguments to these method calls, and
	\item the method call does not return anything or the SUT discards the return value
\end{enumerate*}.
Such method invocation consumes data generated by the SUT instead of producing new data for it.
Therefore, we refer to such method calls as \emph{data consumers}.
These method invocations capture the internal computation results of the SUT, which are usually unobservable by test assertions.
Therefore, developers verify such method invocations with mock assertions to observe such internal computation results.

\begin{figure}[t]
	\centering
	\includegraphics[width=\linewidth]{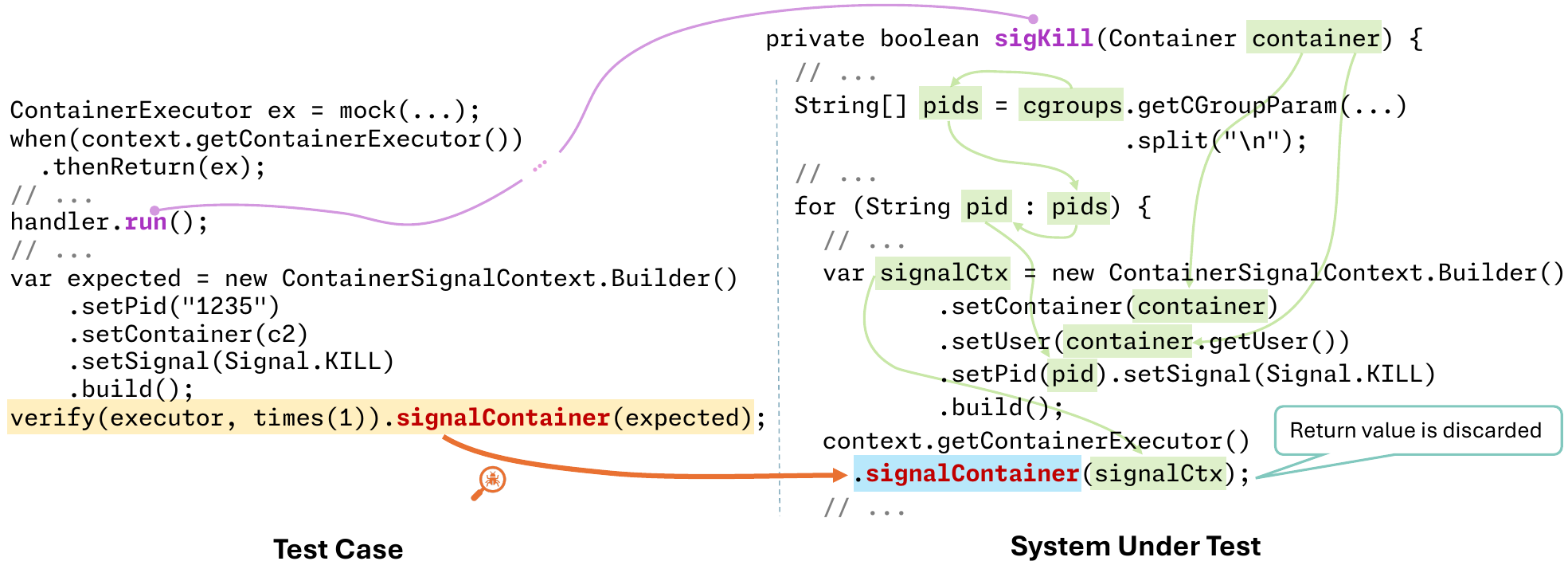}
	\caption{A Mock Assertion Verifying a Method Call Acting as a Data Consumer (Project \textsc{Hadoop}). The method invocations uses data generated within the SUT while its return value is discarded.}
	\label{fig:dataflow-example}
	\vspace{-1em}
\end{figure}

Figure~\ref{fig:dataflow-example} illustrates an example in the \textsc{Hadoop} project.
In the SUT, method \code{sigKill} constructs the argument \code{signalCtx} using results computed from the argument \code{container} and a private field \code{cgroups}.
The data encapsulated in \code{signalCtx} is then passed to the method call \code{signalContainer}, which is responsible for terminating the execution of a container.
Although the method returns a boolean value indicating whether the signal was successfully sent, this return value remains unused by the SUT.
In this example, the method \code{signalContainer} is invoked in a fire-and-forget manner.
Instead of producing new data for the SUT, the method call to \code{signalContainer} consumes data computed internally within the SUT to perform a side effect, namely, sending a signal to the container.
In the corresponding test case, developers verify the method call to \code{signalContainer} using a mock assertion.
Specifically, they check whether the argument passed to \code{signalContainer} matches the expected value to ensure that the SUT sends the correct signal to the container.

Among the 348 sampled method calls, 179 (51\%) act as data consumers in the SUT.
Generally, the SUT passes its internal states as arguments to method calls to perform side effects.
Additionally, it often discards the returned data or simply checks the return value to confirm the success of the operation.
During the verification of such method calls, developers typically capture the arguments to validate that the internal states passed to these calls are correct, ensuring that the intended side effects are achieved.
Among the 179 method calls identified as data consumers, developers check that the arguments match the expected values in 149 (83\%) cases.
This phenomenon shows that verifying the correctness of arguments is a crucial aspect of testing data consumer methods.

\begin{summary}
	Developers use mock assertions to verify method calls that are invoked conditionally (52\%).
	The frequency of such method calls is a proxy of the correctness of the control flow logic in the SUT.
	Furthermore, developers verify method calls acting as data consumers (51\%) to ensure the correctness of internal computation results.
\end{summary}

\section{RQ4: How Do Mock Assertions Complement Test Assertions in Fault Detection?}\label{sec:rq4}
RQ4 aims to investigate the effectiveness of mock assertions in detecting potential faults.
Specifically, we focus on the test cases where both mock assertions and test assertions are utilized.
By examining the interplay between these two types of assertions, we aim to gain insights into how they collectively contribute to the fault detection capabilities of the test cases.

\subsection{Empirical Study Approach}
To achieve this goal, we employ mutation analysis~\cite{DBLP:journals/tse/JiaH11} as a proxy to assess the adequacy of the tests in detecting potential faults.
This technique creates mutants of the SUT by injecting artificial faults and checking if the test cases can ``kill'' them \ie{the tests are expected to fail}.

To understand how mock assertions complement test assertions in detecting potential faults, we performed mutation analysis on the variants of the test cases that use both types of assertions.
Specifically, for each test case \(\mathcal{T}\) that utilizes both types of assertions, we performed mutation analysis on the original tests and three of their variants:
\begin{itemize}
	\item \(\mathcal{T}\) (original): The original test case.
	      It kills the mutants that either type of assertion can kill.
	\item \(\mathcal{T}_{MA}\) (with mock assertions only):
	      It kills the mutants that mock assertions in \(\mathcal{T}\) can kill.
	\item \(\mathcal{T}_{TA}\) (with test assertions only):
	      It kills the mutants that test assertions in \(\mathcal{T}\) can kill.
	\item \(\mathcal{T}_{NA}\) (without any assertion):
	      It kills the mutants that lead to a crash before any assertion in \(\mathcal{T}\) is executed.
\end{itemize}
These variants exercise the same code in the SUT with the same test inputs, with the only difference in the assertions they use.
In this case, we can isolate the contributions of each assertion type in fault detection.
By definition, {\small \(Killed(\mathcal{T}) = Killed(\mathcal{T}_{MA}) \cup Killed(\mathcal{T}_{TA})\)}, and {\small \(\small Killed(\mathcal{T}_{NA}) \subseteq \left(Killed(\mathcal{T}_{MA}) \cap Killed(\mathcal{T}_{TA})\right)\)}.

To construct a dataset of such test cases, we adopted a semi-automatic approach.
First, we executed the test cases to record their method invocations.
Next, we selected the test cases that invoked both mock assertion APIs provided by \mockito{} and assertions APIs provided by assertions frameworks~\eg{\junit{}}.
Out of the 4,652 test cases interacting with test doubles (Table~\ref{tab:subjects}), we identified 1,071 such test cases.
However, it is important to note that invoking both types of assertion APIs does not necessarily mean the test case uses both types of assertions.
This is because some \junit{} assertions are dependent on mock assertions.
As illustrated in Figure~\ref{fig:mock-assertion-illustration}, a \junit{} assertion can predicate a value captured by mock assertions.
In such cases, these \junit{} assert statements are considered part of mock assertions.
Distinguishing such assert statements requires understanding the role of the assert statements.
Therefore, we opted for a manual approach.
Specifically, we randomly sampled 283 test cases out of the 1,071 for manual review.
Such a sample size ensures a confidence level of 95\% and a margin of error of 5\%.
During the manual review, we found that in 43 test cases, all the \junit{} assertions were dependent on mock assertions.
In other words, there were no test assertions in these test cases.
Consequently, they were excluded from the mutation analysis.
For the remaining 240 test cases, we removed the test assertions to create \(\mathcal{T}_{MA}\), removed the mock assertions to create \(\mathcal{T}_{TA}\), and removed all the assertions to create \(\mathcal{T}_{NA}\).
Notably, we kept the inlined method calls when removing test assertions.
For example, \code{assertEquals(..., sut.foo())} was turned into \code{sut.foo()} to preserve the coverage of the test case.

\begin{table}[t]
	\centering
	\caption{Distribution of Mutants Types Covered by Selected Test Cases}\label{tab:mutant-types}
	\smaller
\setlength{\tabcolsep}{3pt}
\begin{tabularx}{\linewidth}{Lr|Lr|Lr}
\toprule
\textbf{Mutation Operatior} &
\textbf{\# Mutants} &
\textbf{Mutation Operator} &
\textbf{\# Mutants} &
\textbf{Mutation Operator} &
\textbf{\# Mutants}
\\
\midrule
{\ttfamily NonVoidMethodCall}      & 2,085    &    {\ttfamily ArgumentPropagation}    & 349   &   {\ttfamily BooleanFalseReturnVals} & 46    \\
{\ttfamily RemoveConditional}      & 1,931    &    {\ttfamily NakedReceiver}          & 203   &   {\ttfamily PrimitiveReturns}       & 45    \\
{\ttfamily NegateConditionals}     & 959      &    {\ttfamily NullReturnVals}         & 199   &   {\ttfamily RemoveSwitch}           & 38    \\
{\ttfamily MemberVariable}         & 844      &    {\ttfamily ConditionalsBoundary}   & 95    &   {\ttfamily Increments}             & 8     \\
{\ttfamily InlineConstant}         & 598      &    {\ttfamily Math}                   & 88    &   {\ttfamily RemoveIncrements}       & 8     \\
{\ttfamily VoidMethodCall}         & 552      &    {\ttfamily EmptyObjectReturnVals}  & 83    &   {\ttfamily Switch}                 & 7     \\
{\ttfamily ConstructorCall}        & 357      &    {\ttfamily BooleanTrueReturnVals}  & 83    &   {\ttfamily InvertNegs}             & 1     \\
\midrule
\(\hookrightarrow\)
& \(\hookrightarrow\)
& \(\hookrightarrow\)
& \(\hookrightarrow\)
& \textbf{Total}       & 8,579 \\
\bottomrule
\end{tabularx}

\end{table}

Subsequently, we ran the PIT mutation testing tool~\cite{Tool:pit} on the variants of these test cases.
We enabled all the builtin mutation operators of PIT to increase the diversity of the injected faults.
For the killed mutants, we attribute them to the two assertion types as follows:
\begin{itemize}
	\item {\small \(\mathcal{M}_{MA} = Killed(\mathcal{T}_{MA}) - Killed(\mathcal{T}_{NA})\)} is the set of mutants killed by mock assertions.
	\item {\small \(\mathcal{M}_{TA} = Killed(\mathcal{T}_{TA}) - Killed(\mathcal{T}_{NA})\)} is the set of mutants killed by test assertions.
	      % \item {\small \(\mathcal{M}_{B} = \mathcal{M}_{MA}\cap \mathcal{M}_{TA}\)} is the set of mutants killed by both types of assertions.
\end{itemize}

During the mutation analysis, PIT seeded 26,481 faults into the SUTs, among which 8,579 were covered by the test cases.
Since all these variants share the same test inputs and execution paths, they cover the same set of mutants.
Table~\ref{tab:mutant-types} show the distribution of the 8,579 covered mutants.
They are produced by 21 mutation operators, with \code{NonVoidMethodCall} being the most frequent and \code{InvertNegs} being the least.
Then, our analysis will focus on these covered mutants to evaluate the effectiveness of different assertion types in detecting injected faults.

\subsection{Empirical Study Results}\label{sec:rq4-results}

\begin{figure}[t]
	\centering
	\includegraphics[width=0.97\linewidth]{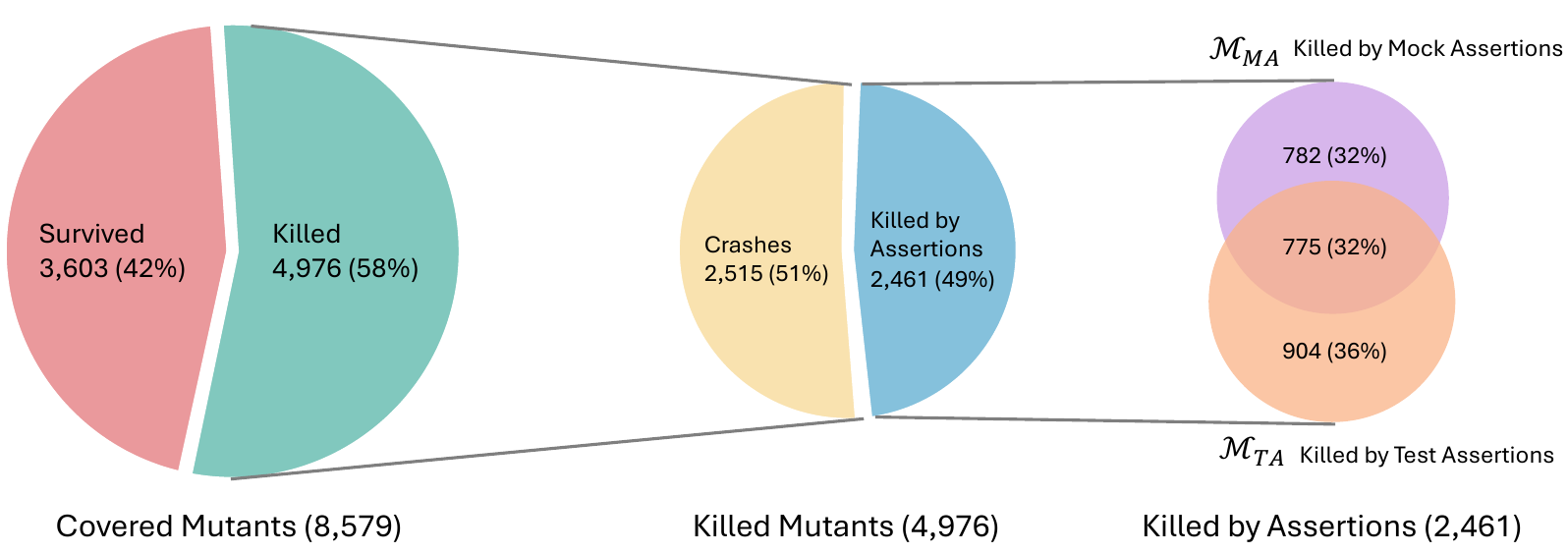}
	\caption{Breakdown of the Covered Mutants by Each of the Assertion Types}
	\label{fig:mutation-analysis}
	% \vspace{-1em}
\end{figure}

Figure~\ref{fig:mutation-analysis} shows the distribution of the 8,579 mutants covered by the 240 test cases.
There are 4,976 mutants killed by at least one of the variants, accounting for 58\% of the covered mutants.
Among the 4,976 killed mutants, 2,515 (51\%) are killed by \(\mathcal{T}_{NA}\).
These mutants lead to crashes when the SUT is running, they can be detected even if no assertion was written.
The remaining 2,461 (49\%) mutants are not killed by \(\mathcal{T}_{NA}\) but killed by any of \(\mathcal{T}_{MA}\) and \(\mathcal{T}_{TA}\).
We find that mock assertions and test assertions are complementary in detecting potential faults.
As shown in Figure~\ref{fig:mutation-analysis}, 68\% of the mutants are killed by only one assertion type.
Notably, 782 of 1,557 (50\%) of the injected faults detected by mock assertions were not detected by test assertions, and the Jaccard similarity between \(\mathcal{M}_{MA}\) and \(\mathcal{M}_{TA}\) is only 32\%.
The result shows that mock assertions can complement test assertions in detecting potential faults.

The complementariness is also found at test case level.
In 109 (45\%) of the 240 test cases, mock assertions and test assertions killed at least one mutant that the other assertion type did not kill.
Moreover, in 138 (58\%) of the 240 test cases, mock assertions killed at least one mutant that test assertions did not kill.
Mock assertions and test assertions complement each other in these test cases.
This shows the importance of mock assertions in assuring software quality.

\begin{figure}[t]
	\centering
	\includegraphics[width=\linewidth]{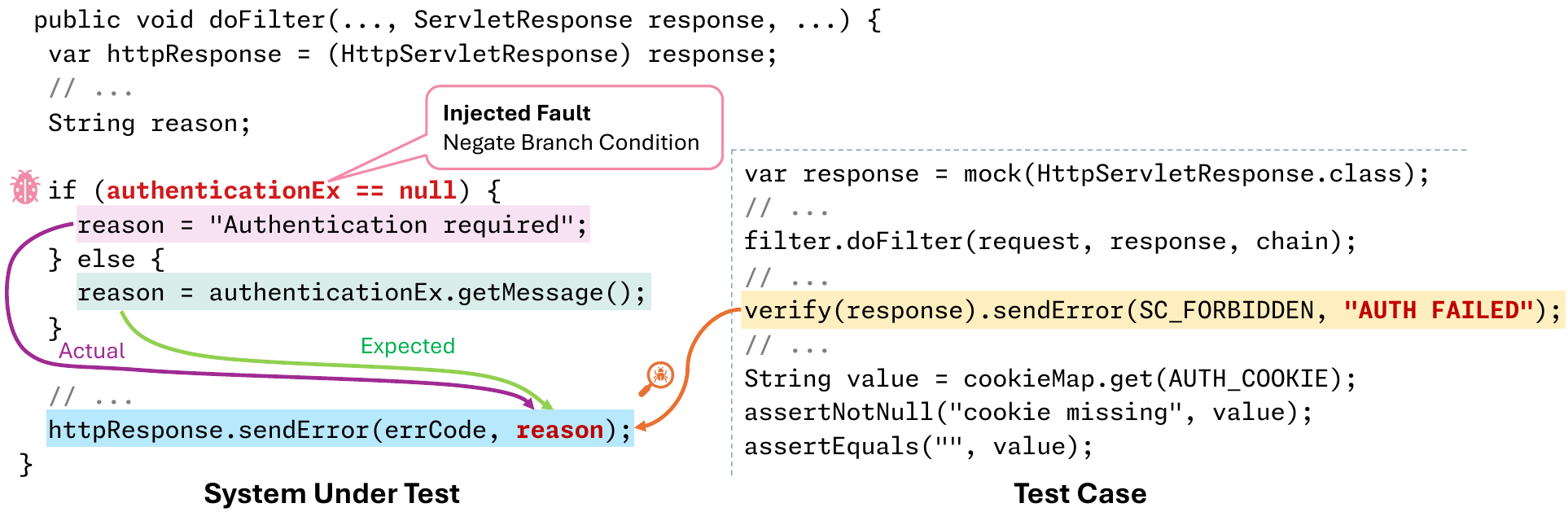}
	\caption{An Incorrect Branch Condition Implementation Detected by a Mock Assertion (Project \textsc{Hadoop}). The method \code{sendError} interacts with external resources. The method invocation is a data consumer in the SUT.}
	\label{fig:ma-detection}
\end{figure}

Figure~\ref{fig:ma-detection} shows an example in project \textsc{Hadoop} where mock assertions detect a fault that test assertions do not detect.
In the SUT, depending on whether \code{authenticationEx} is \code{null}, different strings containing the \code{reason} for authentication failure are sent to the HTTP client via the method \code{sendError}.
This method interacts with \emph{external resources} and it is a candidate to be verified by mock assertions.
In addition, such a method invocation is a \emph{data consumer} that captures an internal computation result.
During mutation analysis, an injected fault negates the condition of the \code{if} statement, which makes the SUT send a different \code{reason} to the client via \code{sendError}.
Since the \code{reason} is sent over the internet and is never returned to the test case, test assertions cannot observe such a difference.
However, mock assertions enable developers to check the arguments passed to the method \code{sendError}.
Specifically, developers expect the second argument to be \code{"AUTH FAILED"}.
During test execution, the mock assertion failed since the second argument for this method call was \code{"Authentication required"}.
In this example, mock assertions complement test assertions by ensuring the correctness of important side effects and internal computation results.

\begin{table}[t]
	\centering
	\caption{Distribution of the Mutants that are Detected by Mock Assertions Only}\label{tab:ma-killed-mutant}
	
\smaller
\setlength{\tabcolsep}{3pt}
\begin{tabularx}{\linewidth}{Lr||Lr}
	\toprule
	\textbf{Method Behavior\textsuperscript{1}}& 
    \textbf{\# Mutants}&
	\textbf{Interactions with SUT\textsuperscript{2}}& 
    \textbf{\# Mutants}\\
	\midrule
    External Resource       & 114 (44\%) & \multirow{2}{*}{Conditional Invocation} & \multirow{2}{*}{198 (77\%)} \\ 
    State Mutator           & 99 (38\%)  & \\  
    Callback                & 41 (16\%)  & \multirow{2}{*}{Data Consumer} & \multirow{2}{*}{115 (45\%)} \\  
    Accessor                & 12 (5\%)   & \\  
    \midrule
    \textbf{Sample Size}    & 258 (CL=95\%, ME=5\%)  && 258 (CL=95\%, ME=5\%) \\
	\bottomrule
\end{tabularx}
\begin{tablenotes}
    \item \scriptsize 1 The overlaps between External Resource and State Mutators is 1, and the overlap between State Mutator and Callback is 7.
    \item \scriptsize 2 The overlap between these two categories is 83.
\end{tablenotes}
	% \vspace{-0.5em}
\end{table}

To get a better understanding of the 782 injected faults that were only detected by mock assertions, we randomly sampled 258 of them for manual analysis, such a sample size ensures a confidence level of 95\% and a margin of error of 5\%.
Specifically, we followed the criteria in RQ2 and RQ3 to categorize the verified method calls of the failing mock assertions.
Table~\ref{tab:ma-killed-mutant} shows the distribution of the 258 mutants.
Among these sampled mutants, 246 (95\%, 114+99+41-1-7) were detected by a mock assertion verifying a method in the category of \emph{external resources}, \emph{state mutator}, and \emph{callback}.
These methods produce important side effects on the test dependencies and external environments.
The incorrect number of these method calls or incorrect arguments passed to these method calls can cause undesired side effects.
Mock assertions help developers ensure these side effects are produced correctly.
In addition, 230 (89\%, 198+115-83) mutants were detected by a mock assertion verifying a method call that is one of or both \emph{conditional invocation} and \emph{data consumer}.
Mock assertions verifying these method invocations ensure the correctness of the control flow logic and internal computation results.
Notably, there are 12 mutants detected by mock assertions that verify an \emph{accessor}, which seems to misalign with our findings in RQ2.
However, we found 10 of them fall into the category of \emph{conditional invocation}.
Although accessors do not produce side effects, the presence or frequencies of their invocations can help developers validate the control flow logic in the SUT.

\begin{summary}
	We observed a minor overlap (32\%) between the faults detected by mock assertions and test assertions.
	Half of the faults detected by mock assertions were not detected by test assertions.
	Mock assertions complement test assertions by ensuring the desired side effects have been produced, validating control flow logic, and checking internal computation results.
\end{summary}

\section{Discussion}

In this section, we summarize our major findings and provide advice for future researchers and developers.
In addition, we discuss the threats affecting the validity of our findings.

\subsection{Implications}

\paragraph{Leverage Mock Assertions for Stronger Test Oracles.}
Our study demonstrates that mock assertions complement test assertions by ensuring the desired side effects have been produced, validating control flow logic, and checking internal computation results.
Developers should utilize mock assertions to validate program behaviors that are unobservable by traditional test assertions.
Furthermore, while automated test generation techniques often aim for higher test coverage through the use of test doubles~\cite{DBLP:journals/ase/TungDLH24,DBLP:conf/icst/ArcuriFJ17,DBLP:conf/kbse/ArcuriFG14,DBLP:conf/sigsoft/ArcuriFG15,DBLP:conf/kbse/TanejaZX10}, future research should explore the generation of mock assertions to create stronger test oracles in the generated tests.

\paragraph{Avoid Aggressive Mocking.}
Despite some automated test generation techniques generating mock assertions~\cite{Tool:agitarone,DBLP:journals/corr/abs-2208-01321}, they adopt a brute-force approach that results in an excessive number of assertions, leading to fragile tests~\cite{DBLP:conf/kbse/ShamshiriJRFMA15}.
Indeed, our findings indicate that such aggressive use of mock assertions does not align with typical developer practices.
Based on our findings, we recommend three strategies for future researchers and developers to pinpoint critical method invocations during test execution, and generate mock assertions to verify them.
\begin{enumerate}
	\item \textbf{Prioritize methods interacting with external resources, methods mutating program states, and callbacks.}
	      In RQ2, we identified three categories of methods that developers commonly verify with mock assertions.
	      \begin{enumerate*}
		      \item methods interacting with external resources, ensure correct interactions with the external environment;
		      \item methods mutating program states, ensure correct program state transitions; and
		      \item callbacks, ensure event handlers are triggered properly.
	      \end{enumerate*}
	      In general, developers may consider verifying such method invocations to ensures the desired side effects have been produced.
	      Future research may explore automatically identifying such method calls, providing candidates for verification through mock assertions.
	\item \textbf{Focus on method invocations that distinguish execution paths.}
	      In RQ3, we found that developers often verify method invocations executed conditionally based on test inputs or SUT states.
	      The presence/absence/frequencies of such method invocations can serve as a proxy for observing the actual execution path.
	      By enforcing the desired execution paths, verifying such method invocations can reveal potential control flow discrepancies.
	      Developers and researchers working on test generation should consider verifying method calls executed in specific branches, especially when the branches are difficult to reach.
	      Additionally, Zhu~\etal{}~\cite{DBLP:journals/tosem/ZhuWTLCWSZS24} noted that such mock assertions can enhance generated stubs.
	\item \textbf{Capture and validate internal computation results.}
	      In RQ3, we found that developers often verify method calls acting as data consumers within the SUT, which helps validate the internal computation results.
	      For test generation techniques, beyond identifying these method calls, it is also important to generate assertions that predicate the arguments for such invocations.
	      Developers are encouraged to capture the key internal computation results for validation.
\end{enumerate}
These strategies can even be used in combination to generate more effective oracles with fewer mock assertions.
For example, a developer can kill two birds with one stone by verifying a method call living in deep nested branches that also interacts with external resources.
Future research can explore mechanisms~\eg{machine learning} to holistically consider these strategies to generate concise yet effective mock assertions.

\paragraph{Relation between Purity and Mock Assertions.}
In RQ2, we identified categories of methods that either have their behavior influenced by external states or produce side effects.
These methods are often considered impure.
Additionally, in RQ4, we discovered that mock assertions are particularly effective in detecting faults related to side effects.
These findings indicate that developers frequently use mock assertions to verify the behavior of impure methods.
Future research should explore the relationship between method purity and the use of mock assertions.

\subsection{Threats to Validity}

\paragraph{External Validity.}
Our empirical findings might not generalize to other \java{} projects.
To mitigate this threat, we selected large-scale, popular \java{} projects that span various application domains.
The quality and diversity of our subjects bolsters our findings generalizing to a wider range of scanarios.
Also, We selected subjects using \mockito{} to create and configure test doubles in this paper.
There are also other frameworks such as \easymock{}~\cite{Tool:easymock}, \powermock{}\footnote{\powermock{} can use \mockito{} as its backend.
}~\cite{Tool:powermock}, and \jmock{}~\cite{Tool:jmockit}.
Developers can have slightly different practices when using these frameworks due to the difference between their functionalities.
Such differences were not covered by our study.
Indeed, as reported by recent studies~\cite{DBLP:journals/ese/SpadiniABB19,DBLP:conf/icse/FazziniCCLKGO22,DBLP:journals/ese/XiaoZWLLWYW24}, \mockito{} is the most frequently adopted framework, and it is used in over 80\% of the projects that uses a mocking framework.
As a result, our findings can be applied to most of the scanrios where mock assertions are used.

In addition, in RQ4, we considered only the test cases where both types of assertions are used.
This is because developer tend to use both assertion types to complement each other~\cite{Web:testsmells}.
However, there may be cases where the two assertion types complement each other in different test cases.
They exercise different code paths in the SUT with different test inputs, which poses challenges in correctly attributing a killed mutant to the assertions (rather than the test inputs).
Therefore, we did not include such test case for analysis, and our results do not reflect the complementariness of the two assertion types in such test cases.
It is an important future work to investigate how mock assertions and test assertions complement each other in different test cases.

\paragraph{Internal Validity.}
As is the case with most empirical studies, there is a potential for errors introduced by manual analysis.
Since the characterization of mock assertions relies on human judgment, there is a risk of misinterpreting or mischaracterizing certain mock assertions.
These errors could lead to inaccurate characterizations of mock assertions.
To mitigate this threat, the authors cross-checked the results produced by manual analysis and discussed their findings during meetings.
Additionally, we have released our experimental data for readers to validate the results.

During data collection we inferred the SUT following the pattern used by the \junit{} test runner and related work~\cite{DBLP:conf/kbse/ZhuWWLCSZ20,arxiv:tufano2020unit}.
However, this strategy might not accurately identify the SUT when developers do not follow the naming convention.
To mitigate this threat, we manually reviewed a subset of the the test cases in our subjects where this naming convention was not followed.
We found only 5\% of the cases were false negatives, the impact of this threat is minor for our study.

\paragraph{Construct Validity.}
We used mutation analysis as a proxy to assess the fault detection capabilities of two assertion types in RQ4.
However, the mutants generated by PIT~\cite{Tool:pit} might not accurately represent the types of bugs encountered in real-world software development.
To mitigate this threat, we enabled all the mutation operators in PIT to maximize the diversity of injected faults.
Nevertheless, evaluating the mock assertions using real-world bugs is an important future work.

\section{Related Work}
Test doubles enable developers to simulate test dependencies in controlled ways, making it an important technique to support unit testing.
In recent years, several studies (discussed below) have explored the usage of test doubles and developed techniques to facilitate their use.
In this section, we discuss the most relevant studies that focus on the applications of test doubles.

\paragraph{Empirical Studies in Test Doubles}
Several studies have investigated the usage of test doubles in software testing.
Marri \etal~\cite{DBLP:conf/icse/MarriXTHS09} examined the benefits of test doubles in testing file-system-dependent software, highlighting their ability to ease the unit testing process.
They emphasized the need for automated identification of APIs requiring mocking.
Zhu \etal{}~\cite{DBLP:conf/kbse/ZhuWWLCSZ20} addressed this need by identifying code-level characteristics for mocking decisions and developing \mocksniffer{}, a machine learning-based technique for recommending mocking decisions to developers.
Mostafa and Wang~\cite{DBLP:conf/qsic/MostafaW14} analyzed the usage of mocking frameworks in a vast number of open-source \java{} projects, revealing that while mock objects are widely used, only a subset of test dependencies are mocked.
Spadini \etal~\cite{DBLP:conf/msr/SpadiniABB17} explored developers' mocking decisions and found that classes contributing to testing difficulties are often mocked.
They further investigated the evolution of mocking framework usage \cite{DBLP:journals/ese/SpadiniABB19}, highlighting the frequent evolution of API usage related to mock assertions.
Fazzini \etal{}~\cite{DBLP:conf/icse/FazziniCCLKGO22} specifically studied the usage of test doubles in \android{} testing and identified the potential issues they introduce.

These empirical studies offer insights into developers' practices in using test doubles.
They provide statistical evidence on the significance of test doubles and identify challenges that motivate further exploration.
On the same theme, our study delves into the usage of mock assertions, aiming to provide guidance for developers and support future research with empirical evidence.

\paragraph{Test Double Automation Techniques}
The research community has made efforts to automate the creation and maintenance of test doubles.
Saff \etal{}~\cite{DBLP:conf/paste/SaffE04,DBLP:conf/kbse/SaffAPE05} proposed a technique that generates test doubles by capturing interactions between the SUT and its dependencies during system tests.
Tiwari \etal{}~\cite{DBLP:journals/corr/abs-2208-01321} developed \rick{}, which monitors SUT execution in a production environment and generates mock-based test cases to validate common production usage.
Wang \etal{}~\cite{DBLP:conf/sigsoft/WangXYWW21} introduced an automated refactoring technique that migrates inheritance-based mock objects to mocking frameworks.
Other studies have focused on enhancing specific parts of tests involving mocking.
\automock{}, proposed by Alshahwan \etal{}~\cite{DBLP:conf/dagstuhl/AlshahwanJLFST10}, employs symbolic execution to infer post-conditions that must be met by the return values of stubs.
Fazzini \etal{} developed \moka{}~\cite{DBLP:conf/kbse/FazziniGO20}, which collects and generates reusable mock objects for testing \android{} applications.
Zhu \etal{} proposed \stubcoder{}~\cite{DBLP:journals/tosem/ZhuWTLCWSZS24}  to generate and repair stub code for regression testing purposes.
Similarly, Tung \etal{} proposed \textsc{AUTS}~\cite{DBLP:journals/ase/TungDLH24}, which leverages stub code to generate tests for achieving higher code coverage for C/C++ programs.
On the same theme, Li \etal{} proposed \textsc{ARUS}~\cite{li2024automaticallyremovingunnecessarystubbings} to automatically remove unnecessary stubs from test suites.
Additionally, domain-specific mock object generation techniques have been proposed for file systems~\cite{DBLP:conf/kbse/ArcuriFG14,DBLP:conf/icse/MarriXTHS09}, databases~\cite{DBLP:conf/kbse/TanejaZX10}, and networking~\cite{DBLP:conf/sigsoft/ArcuriFG15,DBLP:conf/apsec/Bhagya0G19,DBLP:journals/software/ZhangMLXTH12,DBLP:conf/ssbse/SeranZA23}.

These techniques aim to improve test efficiency, increase coverage, and enhance maintainability by generating tests with test doubles.
However, none of these techniques effectively identified the method invocations that should be verified by mock assertions.
Our study aims to bridge this gap by offering insights and empirical evidence to guide future research in developing such a technique.

\section{Conclusion and Future Work}

In this paper, we conducted the first empirical study to understand the usage of mock assertions in practice.
By analyzing the usage of mock assertions in \testcases{} test cases in \projects{} open-source \java{} projects, we found that although mock assertions are used by 41\% of the test cases that use test doubles, verifying all the method calls to test doubles with mock assertions is not the state of the practice.
In contrast, developers only verify a small part of method calls.
For the verified method invocations, we identified three categories of methods whose invocations are commonly verified by developers and two types of common interactions between the SUT and the verified method invocations.
Last but not least, we found mock assertions complement test assertions by ensuring the desired side effects have been produced, validating control flow logic, and checking internal computation results.
We hope our findings can provide guidance and empirical evidence for future research in this area.
Developers can also benefit from our findings and make better use of mock assertions when writing tests.

An important future work following this study is to explore automated mechanisms to identify the method invocations that should be verified by mock assertions.
One of the key challenges is to balance between the strength \ie{low false negative} and robustness \ie{low false positive} of the test oracle.
Such a technique can reduce the burden on developers when creating test oracles.
In addition, integrating such identification mechanisms with automated test generation techniques enables them to generate proper mock assertions to strengthen the tests.

\section{Data Availability}
Our experimental data is available at \artifactlink{}~\cite{artifact:zenodo/v1}.
The dataset is licensed under the Creative Commons Attribution 4.0 International License.

\begin{acks}
	We would like to express our appreciation to the anonymous reviewers for their insightful and constructive comments.
	We would also like to thank the developers of our studied projects for their generous contributions, which made this research possible.
	This work is supported by Hong Kong Research Grants Council / General Research Fund (HKSAR RGC/GRF, grant no. \texttt{16205722}), the FRQNT/NSERC NOVA program (grant no. \texttt{2024-NOVA-346499}), and National Natural Science Foundation of China (grant no. \texttt{62372219}).
\end{acks}

\bibliographystyle{ACM-Reference-Format}
\bibliography{bibliography/references,bibliography/links,bibliography/artifact}

\end{document}